\documentclass[aps,prl,preprint,groupedaddress]{revtex4}
\usepackage{graphicx}

\begin{document}

\title{Collective excitations in Hydrogen across the pressure-induced 
transition from molecular to atomic fluid}
\author{I.-M. Ilenkov$^{1}$, T. Bryk$^{1,2}$}
\affiliation{
$^1$Institute for Condensed Matter Physics of the
National Academy of Sciences of Ukraine,\\
1 Svientsitskii Street, UA-79011 Lviv, Ukraine\\
$^2$ Lviv Polytechnic National University, UA-79013 Lviv, Ukraine
}

\date{\today}

\begin{abstract}
Dispersion of collective excitations in fluid Hydrogen along the isothermal line T=2500~K,
including the region of molecular-to-atomic fluid transition, is studied 
by ab initio molecular dynamics (AIMD) simulations. The obtained density dependence of the 
adiabatic and high-frequency speed of sound contains a plateau in the region of the 
molecular-to-atomic fluid transition. We show, that the five-variable thermo-viscoelastic 
model of generalized hydrodynamics for pure molecular H$_2$ and pure atomic (H) fluids is 
able to recover perfectly the AIMD-derived time 
correlation functions and sound eigenvalues nicely agree with the numerically estimated 
sound dispersion. In the region of the molecular-to-atomic fluid transition a dynamic model
of chemical reacting mixture should be applied. We discuss the calculations of 
time correlation functions from molecular/atomic units in the reacting mixture from AIMD 
trajectories.
\end{abstract}

\keywords{fluid Hydrogen, collective excitations, molecular-to-atomic fluid transition,
generalized hydrodynamics, ab initio molecular dynamics}

\maketitle

\section {Introduction}

Thermodynamic properties and dynamics of Hydrogen fluid are of special interest 
because of their role in green energy production.
Many theoretical, experimental and simulation studies were dedicated to the structural and dynamic
properties of Hydrogen fluid at ambient conditions and high pressures
\cite{McM12,Pie94,Ala95,Mag96,Bon04,Cel05,Mor10,Che13,Gua15,Bel19,Bry20}. 
Collective dynamics in Hydrogen fluids even in pure molecular state 
is much less studied than this is for simple liquids or water. It is not known how
the extended hydrodynamic modes behave outside the hydrodynamic regime.
{\it Ab initio} studies of collective dynamics in molecular Hydrogen are even
more complicated than this is for simple atomic Hydrogen fluids. The vibrational spectrum
of molecular Hydrogen contains high-frequency intramolecular modes which contribute 
to all time correlation functions, calculated in {\it ab initio} simulations. Recently 
we applied a generalized hydrodynamic approach, in particular the thermo-viscoelastic (TVE) 
dynamic model, to the case of molecular H$_2$ fluid \cite{Ile25} for analysis of 
different types of time correlation functions.

The well-known fact that the increase of applied pressure to Hydrogen fluid causes a 
molecular-to-atomic fluid transition and further metallization of atomic Hydrogen 
fluid puts a question - how this transition is reflected in collective dynamics and 
in dispersion of collective excitations of Hydrogen fluid? Dynamic properties of reacting 
liquids (like H$_2 \to$ H+H in case of Hydrogen) in hydrodynamic approximation (when 
one treats only 
dynamic processes corresponding to fluctuations of conserved quantities) were reported 
in a textbook \cite{Ber} and several theoretical studies \cite{Ber68,Lek72,Lek74,Mad75}. 
The main difference
with the hydrodynamics of one-component simple fluids is in emergence of another 
(in addition to temperature relaxation) relaxing collective mode, 
caused by fluctuations of concentration of molecular/atomic species. This makes 
the hydrodynamic treatment of reacting H$_2 \to$ H+H fluid quite similar to the 
long-wavelength dynamics of a binary liquid \cite{Coh71,Bha74}.
Outside the hydrodynamic regime (where only hydrodynamic modes with lifetime 
$\sim k^{-2}$ contribute to the time correlation functions) non-hydrodynamic collective
modes like stress relaxation, structural relaxation, heat current relaxation 
etc \cite{Bry11} contribute
to the shape of time correlation functions and reflect the effect of atomistic structure 
on transport properties and collective excitations on mesoscales.
However, so far we were not able to find theoretical or combined analytical/simulation studies 
of collective dynamics of reacting H$_2 \to$ H+H fluid outside the hydrodynamic regime.

Therefore our aim in this study is in estimation of the dispersion of longitudinal collective 
excitations in Hydrogen in a wide range of densities, which covers the molecular-to-atomic fluid 
transition, calculations of the spedd of sound as a function of density and an application of a 
method of Generalized Collective Modes (GCM) \cite{deS88,Mry95,Bry01} for analysis of collective 
time correlation functions outside the hydrodynamic regime.
The remaining paper has the following structure: in the next Section we provide details of 
our {\it ab initio} simulations of the Hydrogen fluid in a wide range of densities and 
our theoretical expressions for calculation of the speed of sound and GCM-analysis of time
correlation functions. In Section 3 we report our results on the dispersion of collective 
excitations and suggest an approach how to separate molecular and atomic units from AIMD 
trajectories.  The last Section will contain conclusions of this study.

\section {Ab initio simulations}

We performed {\it ab initio} simulations of molecular Hydrogen fluid at temperature 2500~K 
and six densities in the range 284.73-960.976 kg/m$^3$,
using a system of 1000 particles with periodic boundary conditions. The high temperature 
allowed us to use classical equations of motion for particles, while the electron subsystem 
was brought to the ground state within the density functional theory (DFT) with exchange-correlations
treated in the generalized gradient approximation (Perdew-Burke-Ernzerhof version) 
\cite{Per96}
The time step in {\it ab initio} simulations was 0.2~fs.  At each density after an initial 
equilibration over 2-4 ps we performed the production run over at least  28000 timesteps.

We made use of the electron-ion interaction represented by the projector-augmented waves 
(PAW) potentials \cite{Blo94,Kre99} as implemented in the VASP package. The wave functions 
were expanded in plane waves with the default cut-off energy of VASP potentials. For the
construction of electron density calculations we used only the $\Gamma$ point in the Brillouin 
zone. 
Sixty five wave numbers $k$ were sampled in the calculation of the static and time
correlation functions. The smallest wave numbers was $k_{min}=0.3491$\AA$^{-1}$ at the lowest 
density and $k_{min}=0.5236$\AA$^{-1}$ for the most dense studied system, which actually 
corresponded to the high-pressure metallic liquid \cite{Bry20}.
The calculated $k$-dependent static and time correlation functions were averaged over all 
possible directions of wave vectors having the same absolute value.

During the production run we sampled spatial Fourier components of number density
\begin{equation} \label{dynhyd1}
n(k,t)=\frac{1}{\sqrt{N}}\sum_{j=1}^{N}e^{-i{\bf kr}_j}~,
\end{equation}
longitudinal (L) mass-current density
\begin{equation} \label{dynhyd2}
J^{\scriptscriptstyle L}(k,t)=\frac{m}{\sqrt{N}}\sum_{j=1}^{N}
\frac{({\bf kv}_j)}{k}e^{-i{\bf kr}_j}~,
\end{equation}
and first time derivative of ${\dot J}^L(k,t)$ 
\begin{equation}\label{dJLk}
\frac{dJ^L(k,t)}{dt}= \frac{1}{\sqrt{N}}\sum_{j=1}^{N}
\frac{({\bf kF}_j)-im({\bf kv}_j)^2}{k}e^{-i{\bf kr}_j(t)}~,
\end{equation}
where $m$ is the atomic mass of Hydrogen, ${\bf r}_j(t)$, ${\bf v}_j(t)$ and ${\bf F}_j(t)$
are the coordinate, velocity and force acting on the $j$-th particle.
As it was discussed in \cite{Bry13,Bry23} the spatial Fourier components of the 
energy density are extremely time consuming to sample in DFT in comparison with classical 
MD simulations with effective interatomic potentials. Therefore, for the generalized 
hydrodynamic analysis of collective dynamics in connection with {\it ab initio} simulations 
we made use of the GCM approach for one-component liquids suggested in Ref.\cite{Bry13}.

The longitudinal current-current time correlation functions $F_{JJ}^L(k,t)$, calculated 
directly from the AIMD trajectories and particle velocities along the trajectories were 
then numerically Fourier-transformed to obtain
the longitudinal current spectral functions $C^L(k,\omega)$. The observed maximum location
$\omega_{peak}$ of $C^L(k,\omega)$ at each sampled wave number $k$ were collected into 
numerically-estimated dispersion of collective excitation $\omega(k)$ for each density of 
the studied Hydrogen fluid.

Another way of estimation of the dispersion of collective excitations is via the 
complex eigenvalues of the generalized hydrodynamic matrix \cite{deS88,Mry95}.
The general scheme of the GCM analysis consists in the estimation of the generalized 
hydrodynamic matrix ${\bf T}(k)$ on a chosen basis set of dynamic variables and 
calculations of its matrix elements $T_{ij}(k)$. Then one needs to find
the eigenvalues and eigenvectors of the constructed generalized  
hydrodynamic matrix ${\bf T}(k)$. The pairs of estimated complex-conjugated
eigenvalues correspond to propagating modes, while purely real
eigenvalues - to non-propagating relaxation processes. The associated eigenvectors
allow to calculate so-called mode strengths (weights) of the dynamic
eigenmodes in relevant time correlation functions or in the dynamic
structure factors. 
In this study we initially applied the thermo-viscoelastic five-variable dynamic model of 
generalized hydrodynamics, which for the case of longitudinal dynamics \cite{deS88,Bry10} 
reads as follows:
\begin{equation} \label{a5}
{\bf A}^{(TVE)}(k,t) = \left\{n(k,t), J^L(k,t), \varepsilon(k,t),
\dot{J}^L(k,t), \dot{\varepsilon}(k,t)\right\},
\end{equation}
where the dynamic variable of energy current is $\dot{\varepsilon}(k,t)$.
The construction of the $5\times 5$ generalized hydrodynamic matrix ${\bf T}(k)$
using the set of dynamic variables (\ref{a5}) is performed in the
following way \cite{Mry95,Bry01}
\begin{equation}\label{tk}
{\bf T}(k)={\bf F}(k,t=0){\bf {\tilde F}}^{-1}(k,z=0)~,
\end{equation}
where the majority of matrix elements of $5\times 5$ matrix of static correlation 
functions ${\bf F}(k,t=0)$
and of matrix of Laplace-transformed time correlation functions in Markovian approximation
${\bf {\tilde F}}(k,z=0)$ are taken directly from AIMD. The matrix elements, which need
the knowledge of $\varepsilon(k,t)$ and ${\dot \varepsilon}(k,t)$ are taken as fitting
parameters to recover a number of AIMD-derived time correlation functions by their GCM 
theoretical replicas as it was suggested in \cite{Bry13}. In the original scheme in 
\cite{Bry13} the theoretical GCM replicas were calculated for density-density 
$F_{nn}(k,t)$ and longitudinal current-current $F^L_{JJ}(k,t)$ time correlation functions, 
and the number of fitting parameters was 6 to recover these two AIMD-derived functions.
In this AIMD study, as suggested recently in \cite{Ile25}, we additionally will consider 
another time correlation function $F_{nJ}(k,t)$, 
imaginary part of which upon time Fourier transformation results in the dynamic 
susceptibility $\chi(k,\omega)$, which is related to the dynamic structure factor as
follows
$$
Im \chi(k,\omega)=\frac{m\omega}{k}S(k,\omega)~.
$$ 
All the theoretical GCM replicas of time correlation functions between dynamic variables
from the TVE set (\ref{a5}) are represented  via separable contributions from 
dynamic eigenmodes \cite{Mry95,Bry01b}
\begin{equation}\label{replica}
F_{ij}(k,t)=\sum_{\alpha=1}^{N_v}G_{ij}(k)e^{-z_{\alpha}(k)t}~,
\end{equation}
where $z_{\alpha}(k)$ is the $\alpha$-th real (relaxing mode) or 
complex (propagating) eigenmode, and the weight coefficients of each mode 
contribution $G_{ij}(k)$ is expressed via associated eigenvectors \cite{Mry95}.
We will use the expression (\ref{replica}) for comparison of the AIMD derived time
correlation functions with their GCM replicas.

For the analysis of collective dynamics the density dependence of
adiabatic speed of sound $c_s$ is needed. This quantity is a characteristic
of sound propagation in macroscopic hydrodynamic regime. Another quantity, the high-frequency
speed of sound $c_{\infty}$, reflects the elastic mechanism of sound propagation.
The macroscopic adiabatic speed of sound was calculated by the methodology suggested in 
\cite{Bry23b} as
\begin{equation} \label{cs}
 c_s=\sqrt{c_{\infty}^2-\psi^L(0)/\rho}~,
\end{equation}
where $\psi^L(0)$ is the value of the static correlations for diagonal components of stress 
tensor $\psi^L(0)=V\langle \bar{\sigma}_{zz}\bar{\sigma}_{zz}\rangle/k_BT$, 
where $\bar{\sigma}_{zz}(t)=
\sigma_{zz}(t)-P$  is the fluctuating part of the diagonal component of stress tensor and $P$ is 
the pressure, and $V$ - volume of the simulated system, and $k_B$ - Boltzmann constant.
The high-frequency speed of sound $c_{\infty}$ in (\ref{cs})
was estimated from the long-wavelength asymptote of wavenumber-dependent quantity (normalized
second frequency moments of the current spectra function)
\begin{equation} \label{cinf}
c_{\infty}\stackrel{k\to 0}{=}\frac{1}{k}\left[\frac{\langle {\dot J}^L(-k)\dot{J}^L(k)\rangle}
{\langle J^L(-k)J^L(k)\rangle}\right]^{1/2}~.
\end{equation}

\section{Results and discussion}                                                            

\subsection{Density dependence of dispersion of longitudinal collective excitations in 
Hydrogen fluids}

\begin{figure}
\includegraphics[width=0.60\textwidth]{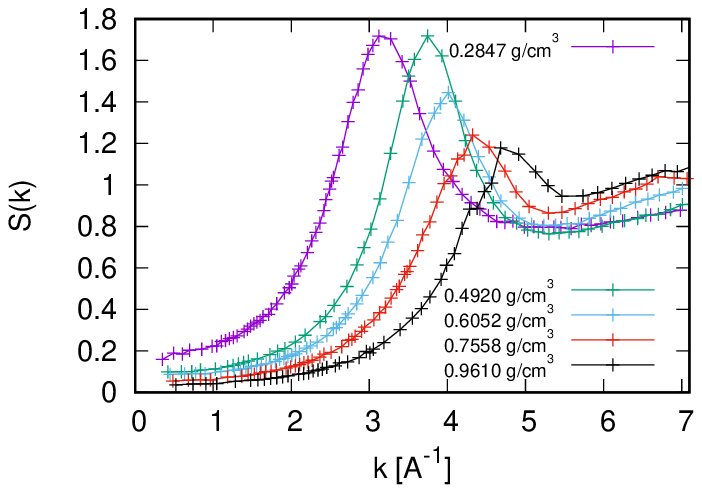}%
\caption{Evolution of the static structure factor $S(k)$ of Hydrogen fluid
with density at temperature 2500~K as obtained directly from AIMD as 
instantaneous density-density correlations.
} \label{str}
\end{figure}

The static structure factors $S(k)$, obtained for several densities of Hydrogen fluid at T=2500~K 
from AIMD in the sampled range of wave numbers as instantaneous 
density-density correlations, are shown in Fig.\ref{str}. The location of main peak of $S(k)$,
which defines the first pseudo-Brillouin zone (Debye wave number) as $k_D=k_p/2$, is 
increasing with density from $k_p=3.12$\AA$^{-1}$ at the lowest studied density to 
$k_p=4.69$\AA$^{-1}$ for the most dense Hydrogen fluid. The amplitude of the main peak of 
$S(k)$ is reducing with the density increase and at the highest density corresponds to the 
pure metallic one-component H fluid with weakly pronounced structural features, as it was 
reported in \cite{Bry20} from pair distribution functions.

\begin{figure}
\includegraphics[width=0.48\textwidth]{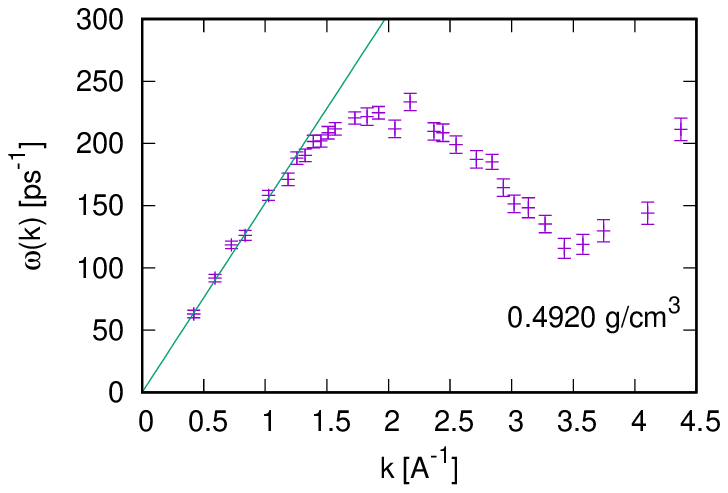}%
\includegraphics[width=0.48\textwidth]{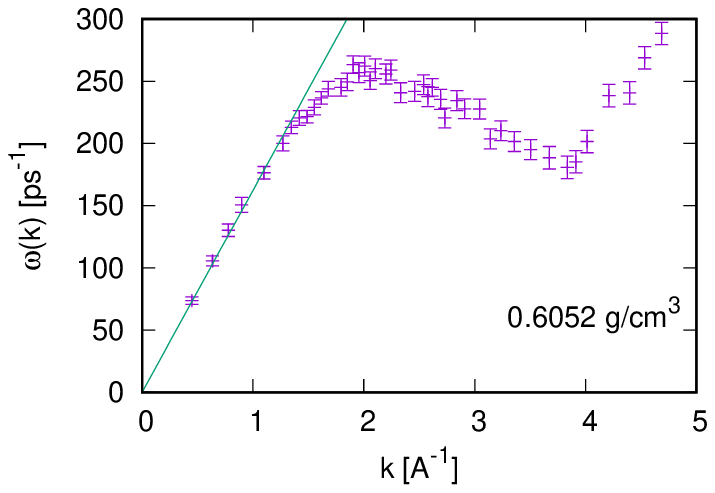}%

\includegraphics[width=0.48\textwidth]{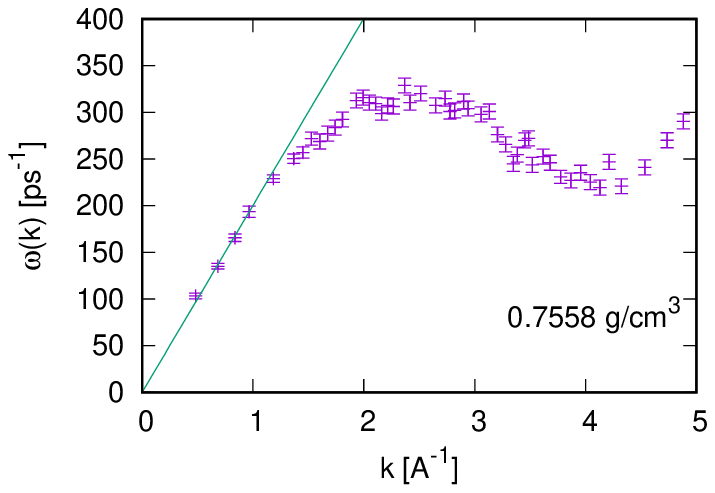}%
\includegraphics[width=0.48\textwidth]{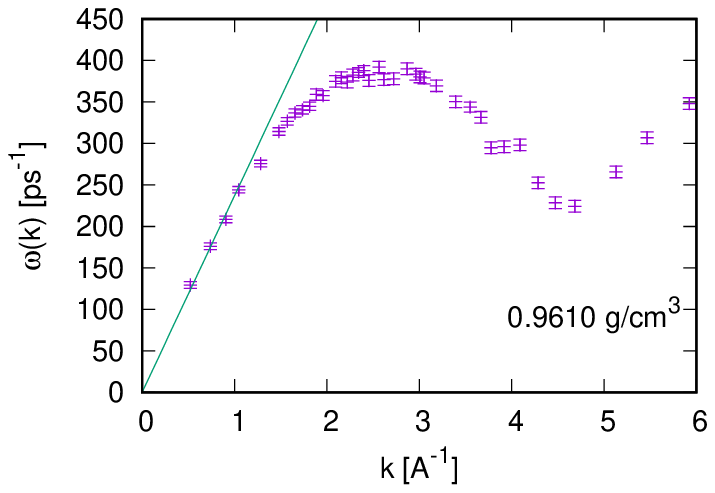}%
\caption{Dispersion of longitudinal collective modes in 
Hydrogen fluid, obtained from peak positions of the longitudinal 
current spectral function $C^L(k,\omega)$, at temperature 2500~K and different densities. 
The straight line corresponds to the hydrodynamic dispersion law $\omega=c_sk$ with 
adiabatic speed of sound $c_s$ obtained from AIMD via Eq.\ref{cs}.
} \label{disp}
\end{figure}

The dispersion of the longitudinal collective excitations at four densities is shown 
in Fig.\ref{disp}. For purely molecular Hydrogen fluid at T=2500~K the dispersion $\omega(k)$ 
was reported recently in \cite{Ile25}. The macroscopic linear dispersion law $\omega_{hyd}(k)=c_sk$
is shown by a green straight line, where the macroscopic adiabatic speed of sound $c_s$ was
calculated from Eq.\ref{cs}. From comparison with the hydrodynamic dispersion law one can estimate,
that in the studied Hydrogen fluids the positive sound dispersion, typical for dense liquids, is
practically absent. 

\begin{figure}
\includegraphics[width=0.60\textwidth]{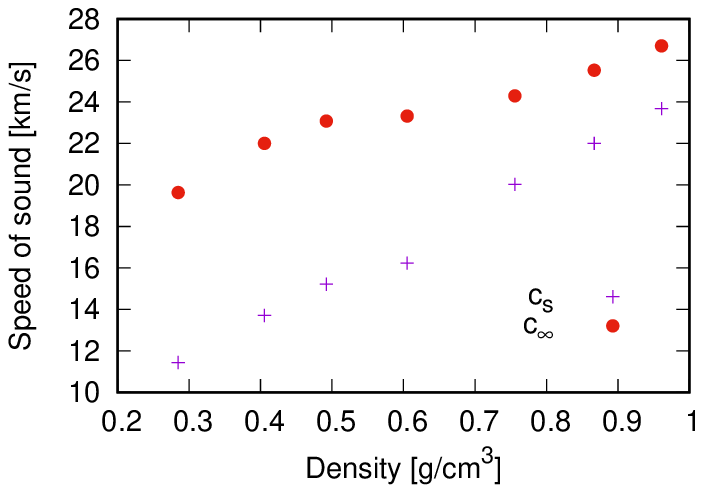}%
\caption{Density dependence of the macroscopic adiabatic $c_s$ and high-frequency
$c_{\infty}$ speed of sound in a Hydrogen fluid at T=2500~K. 
} \label{cscinf}
\end{figure}

The density dependence of the macroscopic adiabatic and high-frequency speeds of sound 
is shown in Fig.\ref{cscinf}. Note, that the adiabatic and high-frequency speeds of sound correspond 
to two different mechanisms of sound propagation in liquids: macroscopic one due to local 
conservation laws and elastic one on mesoscopic length scales due to microscopic forces 
acting on particles. Usually the 
adiabatic speed of sound monotonically increases with density \cite{Bry23b}. However, in the 
region of a structural transformation in liquid (like it was observed in liquid Rb \cite{Bry13b}
and liquid K \cite{Zon21}) or metal-nonmetal transition (like in expanded Hg \cite{Ayr14,Bry25})
the density dependence of $c_s$ can have a plateau or even a minimum at the transition region.
In Fig.\ref{cscinf} one can see a flattering of the adiabatic speed of sound and almost a plateau
in the high-frequency speed $c_{\infty}$ as functions of density at the density where the strong 
deviation in the total charge fluctuations were reported in \cite{Bry20}.

\subsection{Application of the GCM theory to analysis of time-dependent correlations}

\begin{figure}
\includegraphics[width=0.48\textwidth]{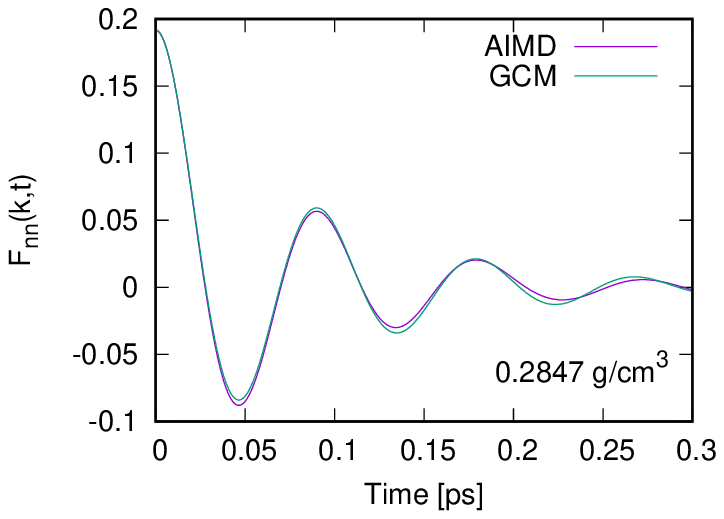}%
\includegraphics[width=0.48\textwidth]{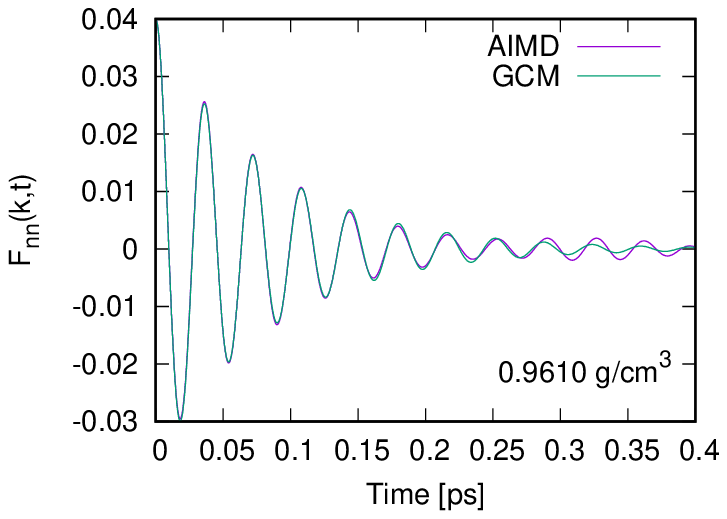}%

\includegraphics[width=0.48\textwidth]{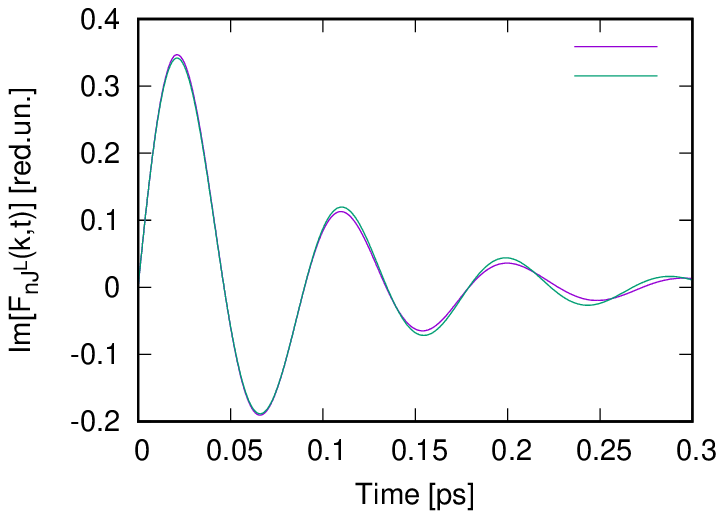}%
\includegraphics[width=0.48\textwidth]{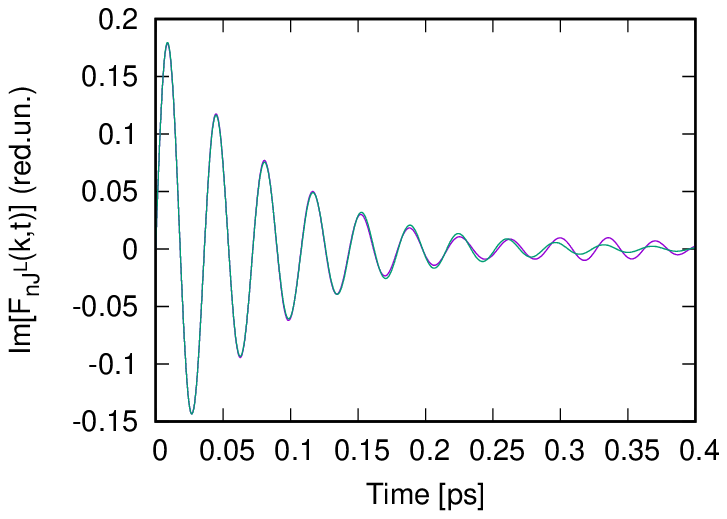}%

\includegraphics[width=0.48\textwidth]{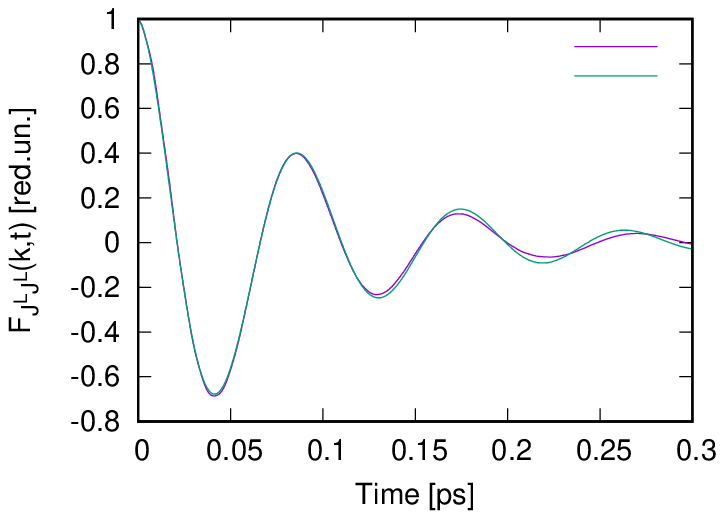}%
\includegraphics[width=0.48\textwidth]{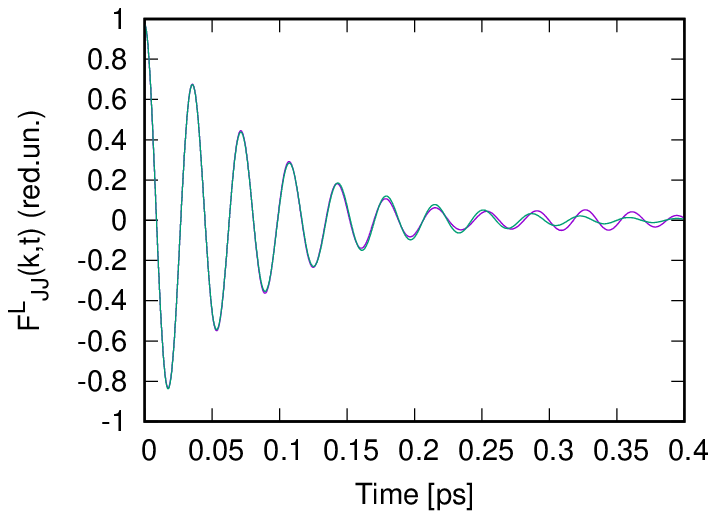}%
\caption{Good recovery of the AIMD-derived time correlation functions
by the 5-variable GCM theory at for pure molecular H$_2$ fluid (density 0.2847 g/cm$^3$
at $k=0.6046$\AA$^{-1}$) and pure metallic H fluid (density 0.9610 g/cm$^3$ 
at $k=0.7405$\AA$^{-1}$). $F_{nn}(k,t)$ - density-density 
correlations, $Im F_{nJ^L}(k,t)$ -imaginary part of susceptibility
in time domain, $F_{J^LJ^L}(k,t)$ - longitudinal current-current correlations.
} \label{tcf}
\end{figure}

Our application of the GCM methodology within the five-variable TVE dynamic model makes 
evidence that
for pure molecular and pure atomic Hydrogen fluids this theoretical approach works
very well. In Fig.\ref{tcf} we show, that the GCM representation (\ref{replica}) of time
correlation functions perfectly recovers the AIMD-derived density-density, density-current
and current-current time correlation functions.

\begin{figure}
\includegraphics[width=0.48\textwidth]{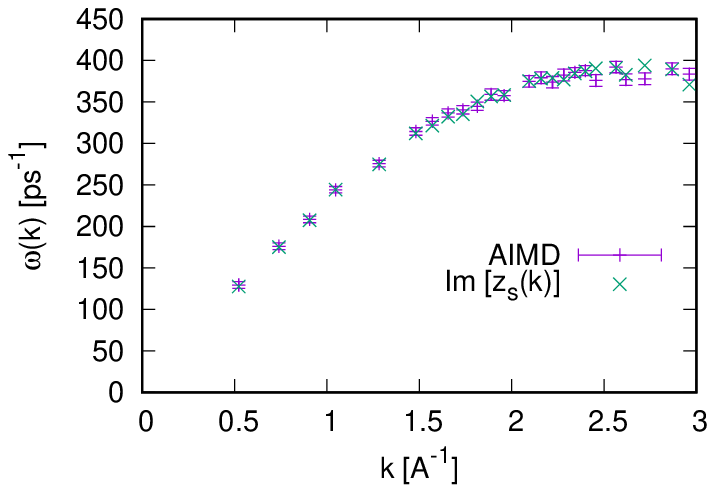}%
\includegraphics[width=0.48\textwidth]{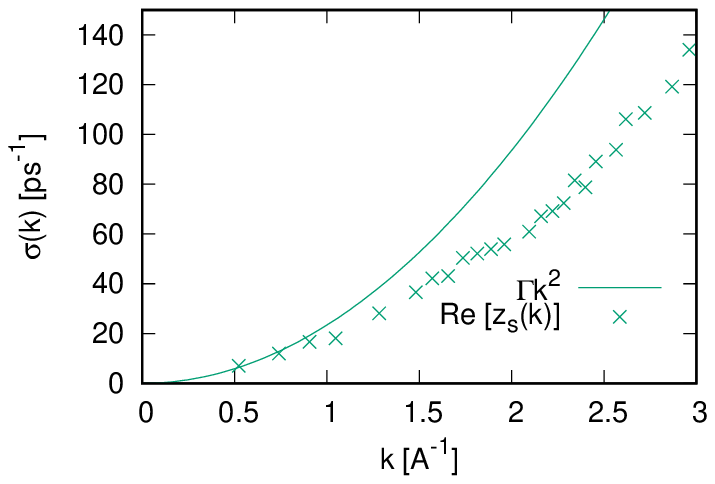}%
\caption{Imaginary (dispersion) and real (damping) parts of complex eigenmodes corresponding to 
the propagating sound modes in atomic Hydrogen fluid at the density 0.9610 g/cm$^3$ obtained 
within the 5-variable GCM theory. 
} \label{eigenval12}
\end{figure}

The good recovery of AIMD-derived time correlation functions should result in the 
same level of precision to describe the corresponding spectral functions in terms 
of partial contributions from different collective eigenmodes. Therefore, it is 
expected that the dispersion of GCM sound eigenmodes should recover the 
dispersion $\omega(k)$ obtained in a purely numerical way via peak positions of 
the current spectral function. Indeed,
the complex-conjugated GCM acoustic eigenvalues, represented in the 
form \cite{deS88,Mry95}
$$
z_{s}(k)=\sigma_s(k)\pm i\omega_s(k)~, 
$$
where the real part corresponds to the $k$-dependent damping, and the imaginary part - 
to the dispersion law. We show the sound eigenvalues in the first pseudo-Brillouin zone 
in Fig.\ref{eigenval12} in comparison with the AIMD-derived peak positions of the longitudinal 
current spectral function $C^L(k,\omega)$ for the highest studied density. 
Also, in Fig.\ref{eigenval12} we show the hydrodynamic behavior of 
dispersion and damping, where the latter should follow the $\Gamma k^2$ dependence, 
and in Fig.\ref{eigenval12}
one can see how the damping corresponds to the hydrodynamic law with the damping coefficient
$\Gamma = 23.4$\AA$^2/ps$. Similar nice recovery of the dispersion $\omega(k)$ is for the
pure molecular Hydrogen fluid (see Ref.\cite{Ile25}).

\begin{figure}
\includegraphics[width=0.50\textwidth]{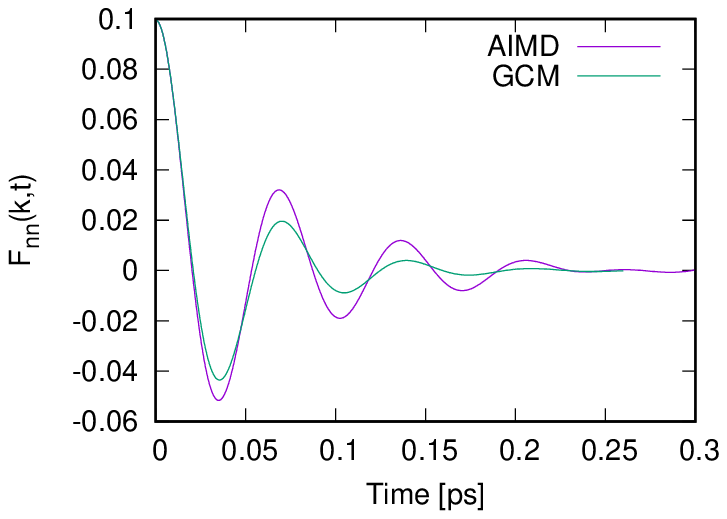}%
\includegraphics[width=0.50\textwidth]{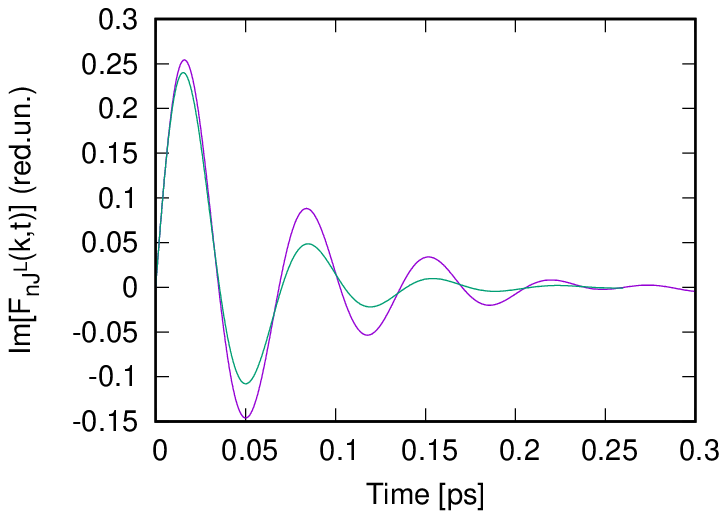}%

\includegraphics[width=0.50\textwidth]{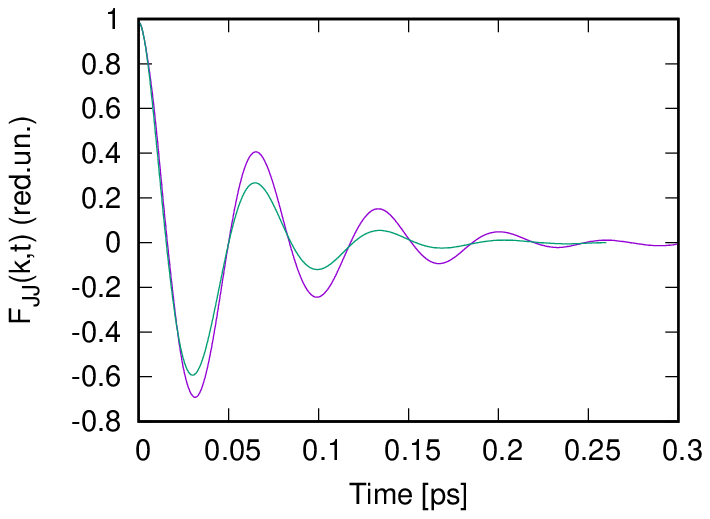}%
\caption{Unsatisfactory recovery of the AIMD-derived time correlation functions             
of a Hydrogen fluid at the density 0.4920 g/cm$^3$ (at 
wave number $k=0.5924$\AA$^{-1}$)  in the region of the molecular-to-atomic fluid
transition by the 5-variable GCM theory. 
} \label{tcf15}
\end{figure}

However, in-between the lowest (purely molecular) and highest (purely atomic) densities
the five-variable TVE dynamic model was unable to recover the AIMD-derived time correlation
functions. In Fig.\ref{tcf15} we show three AIMD-derived time correlation functions 
and their GCM replicas. This means that there are some relaxation processes in the 
region of 
molecular-to-atomic fluid transition which cannot be described by the TVE dynamic model.
Indeed, according to the hydrodynamic theory of reacting fluids \cite{Ber,Mad75} such 
systems have another relaxing mode, which should reflect relaxation of local concentrations
(molecular and atomic units in case of H$_2 \to$ H+H dissociation in our case of the 
Hydrogen fluid). In this case one needs to study time correlation functions as in 
 binary liquids.

\subsection{Separating molecular and atomic units from AIMD in reacting Hydrogen fluids}

\begin{figure}
\includegraphics[width=0.60\textwidth]{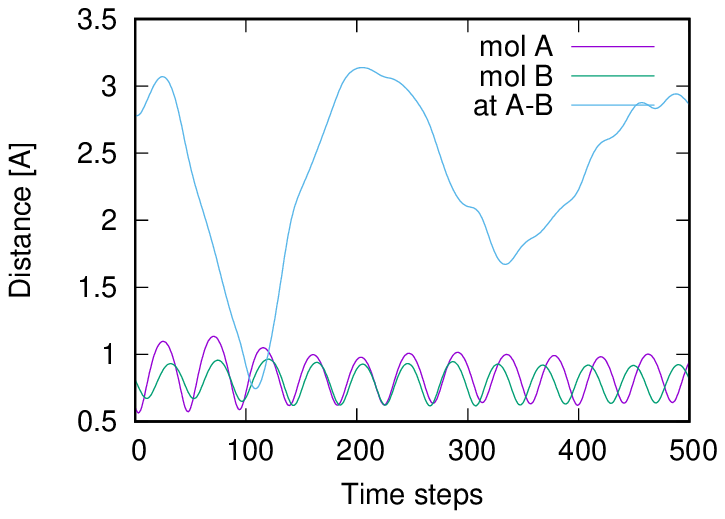}%
\caption{Running distance between Hydrogens of the same (A-A or B-B) and different (A-B) 
molecules showing the collision of two molecules.
} \label{collis}
\end{figure}

In order to enable the GCM description of binary liquids in the case of reacting
 H$_2 \to$ H+H fluid outside the hydrodynamic regime one has to sample dynamic variables 
of partial densities of molecular and atomic components. This is not a simple task with 
AIMD. One has to discriminate between simple molecular collisions and break-up of 
molecules. In Fig.\ref{collis} we show, how the two stable H$_2$ molecules (A-A and 
B-B) collide, that results in a short period of time ($\sim$ 60-70 time steps) when 
the distance between two 
Hydrogens of different molecules (A-B) is comparable with the intramolecular distance.
Hence, in order to separate the molecular and atomic units at each time step of AIMD 
we checked the distances between all the pairs of Hydrogens. Those, which remained 
within the distance $R_M$ for more than consecutive 70 time steps were assigned to be 
molecular units, and others - to be the atomic ones. And this selection of molecular/atomic
units was performed along all the length of trajectories. 

\begin{figure}
\includegraphics[width=0.48\textwidth]{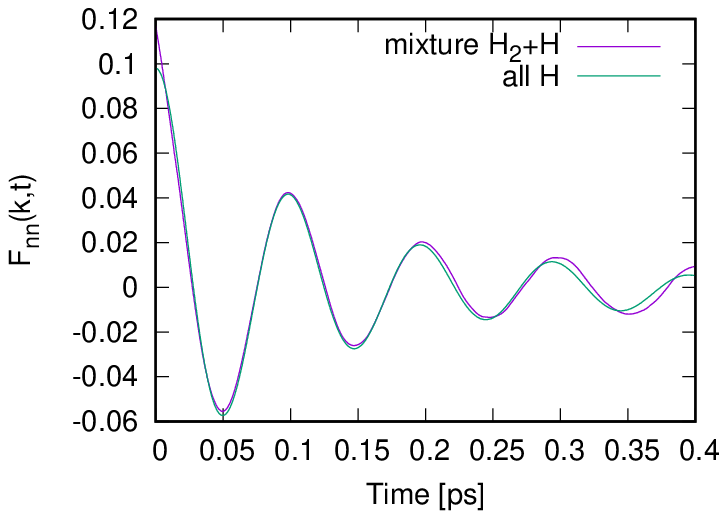}%
\includegraphics[width=0.48\textwidth]{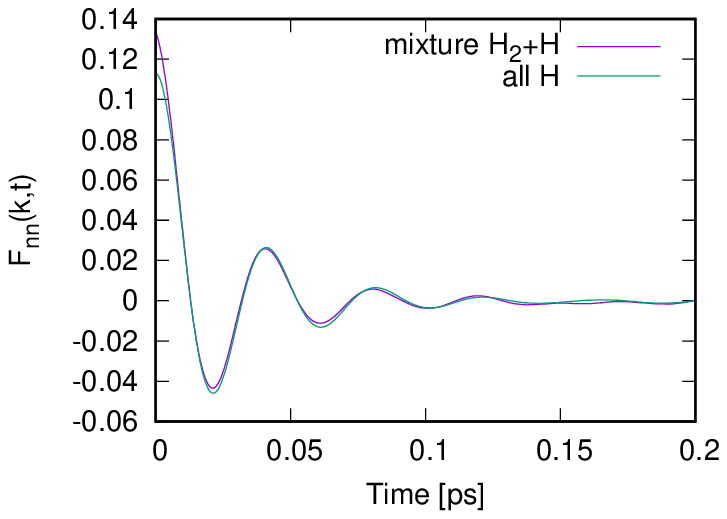}%

\includegraphics[width=0.48\textwidth]{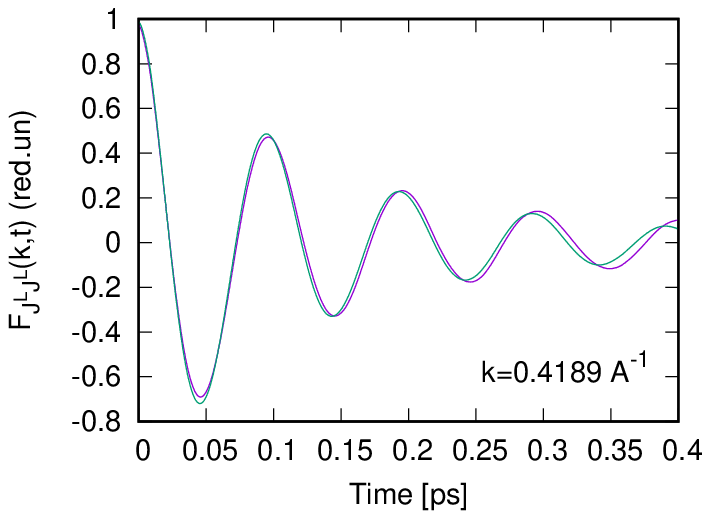}%
\includegraphics[width=0.48\textwidth]{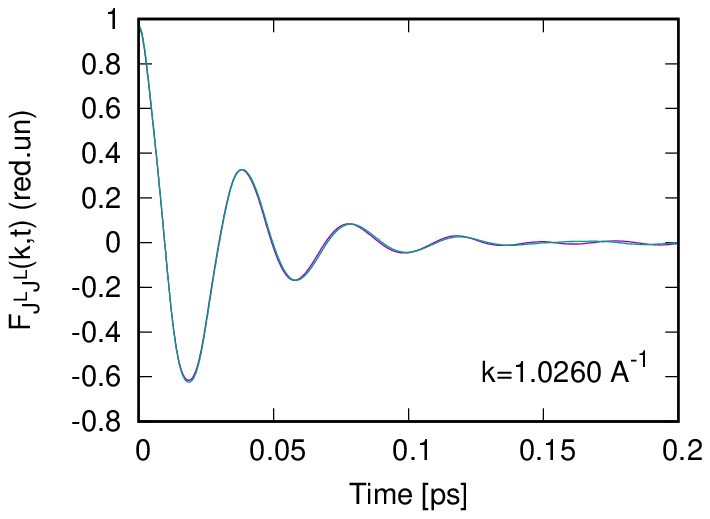}%
\caption{Quality of the decomposition into molecular and atomic units: a comparison 
of the density-density and longitudinal current-current 
time corelation functions, composed from a mixture of molecular and atomic units
(purple solid lines) and from standard estimation using all Hydrogens 
via Eqs. (\ref{dynhyd1}) and (\ref{dynhyd2}) (green line) 
at density 0.4920 g/cm$^3$ and at two wave numbers.
} \label{H2H}
\end{figure}

A correction was needed to this scheme to avoid a sharp change of the number of  
molecular/atomic units. Therefore we made use of a smeared step function of the 
intramolecular distance 
\begin{equation} \label{weight}
w(r_{12})=0.5[1-tanh(\frac{r_{12}-R_0}{d})]~,
\end{equation}
which allows to assign to a pair of particles the molecular weight $w(r_{12})$ 
and atomic weight $1-w(r_{12})$. One can see that the largest intramolecular distance 
in Fig.\ref{collis} is $\sim 1.1$\AA, and the step function $w(r_12)$ in 
Fig.\ref{weight} assigns to this pair of particles the weight practically equal 1.
In case of the break-up of the molecule the distance $r_{12}$ for this pair is increasing
giving the desrease of the molecular weight and increase of the atomic one. At further 
increase of the distance, when $r_{12}>1.3$\AA~ these two particles are purely atomic ones.
\begin{figure}
\includegraphics[width=0.60\textwidth]{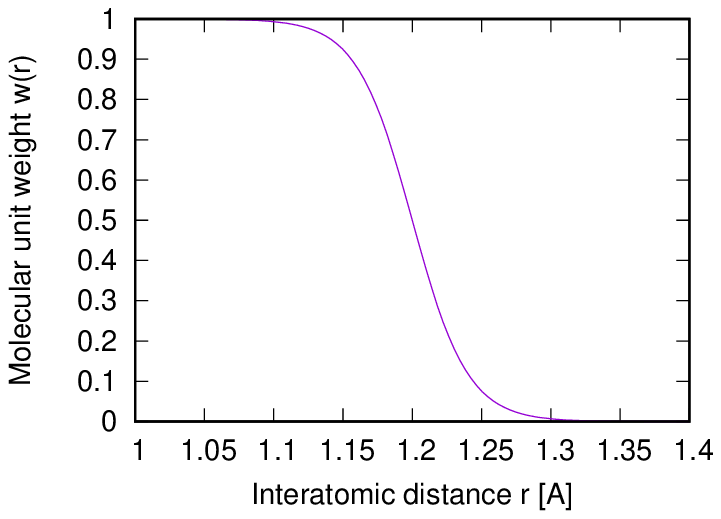}%
\caption{Weight function $w(r)$ for molecular units as a function of intramolecular 
distance between protons, Eq.\ref{weight}, with parameters $R_0=1.2$\AA~ and $d=0.04$\AA .
} \label{wr12}
\end{figure}
Hence, the partial density of molecular units is expressed via the sum over 
all pairs of particles (which were at least 70 consecutive time steps within 
the distance $R_M=1.3$\AA
\begin{equation} \label{nmol}
n_{mol}(k,t)=
\frac{1}{\sqrt{N}}\sum_{(l,m)}w(r_{lm})[e^{-i{\bf kr}_l}+e^{-i{\bf kr}_m}]~,
\end{equation} 
while the patial density of atomic units is 
\begin{equation} \label{nat}
n_{at}(k,t)=
\frac{1}{\sqrt{N}}\sum_{(l,m)}[1-w(r_{lm})][e^{-i{\bf kr}_l}+e^{-i{\bf kr}_m}]
+
\frac{1}{\sqrt{N}}\sum_{j}e^{-i{\bf kr}_j}~.
\end{equation} 
Here, the second term in the right hand side of (\ref{nat}) corresponds to the sum over
atomic particles which do not have neighbors within the distance $R_M=1.3$\AA.
One can make sure that the sum $n_{mol}(k,t)+n_at(k,t)$ results in the same density of 
all Hydrogen particles $n(k,t)$ in Eq.(\ref{dynhyd1}).
We performed calculations to the partial density-density and current-current time 
correlation functions in this molecular/atomic partial representation
and composed from them the total density-density and current-current time correlation 
functions. The comparison with the density-density and current-current time correlation
functions based on dynamic variables (\ref{dynhyd1}) and (\ref{dynhyd2}) shown in 
Fig.\ref{H2H} makes evidence of the correct decomposition.
Based on these partial dynamic variables in molecular/atomic representation one can make 
use of the eight-variable GCM theory for binary liquids \cite{Bry23} for GCM analysis 
of eigenmodes of Hydrogen fluid in the region of molecular-to-atomic fluid transition.
Results of this GCM study will be reported elsewhere.

\section{Conclusions}

We performed {\it ab initio} molecular dynamics simulations for Hydrogen fluid at T=2500~K
and in the range of densities covering the molecular-to-atomic fluid transition.
Our conclusions of this study are as follows:\\
i. We successfully applied the five-variable thermo-viscoelastic dynamic model to 
analysis of AIMD-derived 
time correlation functions and proved that the GCM eigenvalues correctly recover 
the dispersion of acoustic modes for pure molecular and pure atomic (metallic) Hydrogen 
fluid;\\
ii. The estimated density dependence of the adiabatic $c_s$ and high-frequency $c_{\infty}$
 speed of sound show a flattering (and almost plateau for $c_{\infty}$) in the region 
of the molecular-to-atomic transition;\\
iii. We propose a methodology how to calculate partial densities of molecular and atomic
units from AIMD data, which enable the analysis of dynamics of reacting fluids via GCM
methodology for binary liquids \cite{Bry23}.

{\it Acknowledgments}
I.-M.I. was supported by the Project
No. 09/01-2024 from the National Academy of Sciences of Ukraine
for Youth Groups. The  calculations have been performed using the ab-initio total-energy 
and molecular dynamics program VASP (Vienna ab-initio simulation program)
developed at the Institute f\"ur Materialphysik of the Universit\"at Wien
\cite{Kre93,Kre96,Kre96b}.


%
\end{document}